\documentclass[aps,prd,nofootinbib,amsmath,amssymb,superscriptaddress,onecolumn,11pt]{revtex4}
\usepackage{txfonts}
\usepackage{graphicx}
\usepackage{dcolumn}
\usepackage{bm}
\usepackage{amssymb}
\usepackage{latexsym}
\usepackage{booktabs}
\usepackage{amsmath}
\usepackage{multirow}
\usepackage[colorlinks=true, linkcolor=red, citecolor=blue]{hyperref}

\newcommand{\be}{\begin{equation}}
\newcommand{\ee}{\end{equation}}
\newcommand{\bq}{\begin{eqnarray}}
\newcommand{\eq}{\end{eqnarray}}

\begin{document}

\title{Cosmological constraints on neutrinos after BICEP2}

\author{Jing-Fei Zhang}
\affiliation{Department of Physics, College of Sciences, Northeastern University, Shenyang
110004, China}
\author{Yun-He Li}
\affiliation{Department of Physics, College of Sciences, Northeastern University, Shenyang
110004, China}
\author{Xin Zhang\footnote{Corresponding author}}
\email{zhangxin@mail.neu.edu.cn} \affiliation{Department of Physics, College of Sciences,
Northeastern University, Shenyang 110004, China} 
\affiliation{Center for High Energy Physics, Peking University, Beijing 100080, China}

\begin{abstract}
Since the B-mode polarization of the cosmic microwave background (CMB) was detected by the BICEP2 experiment and 
an unexpectedly large tensor-to-scalar ratio, $r=0.20^{+0.07}_{-0.05}$, was found, the 
base standard cosmology should at least be extended to the 7-parameter $\Lambda$CDM+$r$ model. 
In this paper, we consider the extensions to this base $\Lambda$CDM+$r$ model by including additional 
base parameters relevant to neutrinos and/or other neutrino-like relativistic components. 
Four neutrino cosmological models are considered, i.e., the $\Lambda$CDM+$r$+$\sum m_\nu$, $\Lambda$CDM+$r$+$N_{\rm eff}$, $\Lambda$CDM+$r$+$\sum m_\nu$+$N_{\rm eff}$, and $\Lambda$CDM+$r$+$N_{\rm eff}$+$m_{\nu,{\rm sterile}}^{\rm eff}$ models.
We combine the current data, including the Planck temperature data, the WMAP 9-year polarization 
data, the baryon acoustic oscillation data, the Hubble constant direct measurement data, the Planck 
Sunyaev-Zeldovich cluster counts data, the Planck CMB lensing data, the cosmic shear data, and the 
BICEP2 polarization data, to constrain these neutrino cosmological models.
We focus on the constraints on the parameters $\sum m_\nu$, $N_{\rm eff}$, and $m_{\nu,{\rm sterile}}^{\rm eff}$.
We also discuss whether the tension on $r$ between Planck and BICEP2 can be relieved in these 
neutrino cosmological models.
\end{abstract}

 \maketitle

\section{Introduction}
\label{sec:intro}


Recently, the BICEP2 (Background Imaging of Cosmic Extragalactic Polarization) Collaboration reported the detection of the 
B-mode polarization of the cosmic microwave background (CMB), which implies that the primordial gravitational waves (PGWs) 
are likely  to have been detected~\cite{bicep2}. If confirmed by upcoming experiments,  
the BICEP2's result will greatly impact on the fundamental physics.
The tensor-to-scalar ratio derived by the observed B-mode power spectrum is
unexpectedly large, $r=0.20^{+0.07}_{-0.05}$, with $r=0$ disfavored at the 7.0$\sigma$ level~\cite{bicep2}.
This result is in tension with the upper limit $r<0.11$ (95\% CL) deduced from the measurements of temperature power spectrum 
by the Planck Collaboration (Planck+WP+highL, where WP refers to the WMAP 9-year polarization data and highL refers to the 
temperature data from ACT and SPT)~\cite{planck}.
One simple way of relieving this tension is to allow for a negative running of the scalar spectral index of order $10^{-2}$, which challenges 
the design of the inflation models since the usual slow-roll inflation models predict a negligible running (of order $10^{-4}$).

To reduce the tension, more possibilities should be explored. One interesting suggestion is to consider additional sterile neutrino species 
in the universe~\cite{zx14,WHu14}. Since the tensor-to-scalar ratio $r$ is found to be around 0.2, the standard cosmology 
should at least be extended to $\Lambda$CDM+$r$ model (now this is the base model with seven parameters). Thus, the model with sterile 
neutrino is naturally called $\Lambda$CDM+$r$+$\nu_s$ model, in which two additional parameters, $N_{\rm eff}$ and $m_{\nu,{\rm sterile}}^{\rm eff}$, 
are included. It is shown that in the $\Lambda$CDM+$r$+$\nu_s$ model the tension between Planck and BICEP2 can be 
greatly relieved at the expense of the increase of $n_s$~\cite{zx14,WHu14}. Moreover, actually, by including a sterile neutrino species in the universe, 
not only the tension between Planck and BICEP2 is relieved, but also the other tensions between Planck and other astrophysical observations, 
such as the $H_0$ direct measurement, the cluster counts, and the galaxy shear measurement, can all be 
significantly reduced.\footnote{In fact, even before the release of the Planck temperature data, the effects of neutrino mass and additional neutrino species 
in relieving the tension between CMB+BAO and 
other observations, such as $H_0$ and cluster counts, were discussed~\cite{SPT,Burenin2013}. 
Then, after the Planck data release, the result was further confirmed; see, e.g., 
Refs.~\cite{tsz,snu1,snu2,snu3,Gariazzo:2013gua}. }
Thus, the model with sterile neutrino seems to be an economical choice for the cosmology today. 
Furthermore, by combining the Planck+WP with the baryon acoustic oscillations (BAO), $H_0$, Sunyaev-Zeldovich (SZ) cluster counts, CMB lensing, galaxy shear, 
and BICEP2 data, it is found that in the $\Lambda$CDM+$r$+$\nu_s$ model 
the existing cosmological data prefer $\Delta N_{\rm eff}>0$ at the 2.7$\sigma$ level and a nonzero mass of sterile neutrino at the 3.9$\sigma$ level~\cite{zx14}. 
(See also Ref.~\cite{WHu14} for a similar analysis.)

Other proposals to address the large B modes include, e.g., foregrounds or some uncounted temperature-polarization leakage~\cite{Liu:2014mpa}, 
non-standard inflation models or more general early-universe scenarios~\cite{Harigaya:2014qza,Nakayama:2014koa,Brandenberger:2014faa,
Contaldi:2014zua,Miranda:2014wga,Gerbino:2014eqa,McDonald:2014kia,Hazra:2014a,Hazra:2014b}, 
large-field excursions~\cite{Kehagias:2014wza,Lyth:2014yya}, primordial magnetic fields~\cite{Bonvin:2014xia}, 
topological defects~\cite{Lizarraga:2014eaa,Moss:2014cra}, spatial variation of $r$~\cite{Chluba:2014uba}, and so on. 
Obviously, the forthcoming new data from, e.g., Planck and Keck array are expected to improve the foreground model and provide more tight constraints 
on the B modes, resolving the current tension problem.

In this paper, we will consider neutrinos and extra relativistic components within the base $\Lambda$CDM+$r$ model.
We will use the current data to constrain the models with neutrinos. The models we consider in this paper include: (i) the active neutrinos with additional 
parameter $\sum m_{\nu}$, (ii) the extra relativistic components with additional parameter $N_{\rm eff}$, (iii) the active neutrinos along with the 
extra relativistic components with additional parameters $\sum m_{\nu}$ and $N_{\rm eff}$, and (iv) the massive sterile neutrino with 
additional parameters  $N_{\rm eff}$ and $m_{\nu,{\rm sterile}}^{\rm eff}$. The observational data we use in this paper are from Planck+WP+BAO, 
$H_0$ direct measurement, Planck SZ cluster counts, Planck CMB lensing, cosmic shear measurement, and BICEP2. 
This work will provide a detailed cosmological analysis on the models with neutrinos under the consideration of the BICEP2 data.

The paper is organized as follows. In Sec.~\ref{sec:cosmol},  we briefly describe the cosmological models with neutrinos and the observational data. 
In Sec.~\ref{sec:result}, we present the fit results and discuss these results in detail. Conclusion is given in Sec.~\ref{sec:concl}.

\section{Models, parameters, and data}\label{sec:cosmol}

\subsection{Cosmological models involving neutrinos}
The cosmology with neutrinos has been described in detail and reviewed by the WMAP Collaboration~\cite{wmap5,wmap7,wmap9} and the Planck Collaboration~\cite{planck}. 
In this paper, our conventions are consistent with those adopted by the Planck Collaboration~\cite{planck}, i.e., those used in the {\tt camb} Boltzmann code. 
So, we will not describe in detail the equations but only specify the models with different parameters; for the details about the cosmology with neutrinos 
we refer the reader to Refs.~\cite{planck,wmap5,wmap7,wmap9}.

Under the current situation that the large PGWs have been discovered, the base cosmology should be extended to the 7-parameter $\Lambda$CDM+$r$ model. 
The base parameters for this model are:
$$\{\omega_b,~\omega_c,~100\theta_{\rm MC},~\tau,~n_s,~\ln (10^{10}A_s),~r_{0.05}\},$$
where $\omega_b\equiv \Omega_b h^2$ and $\omega_c\equiv \Omega_c h^2$ are the present-day baryon and cold dark matter densities, respectively, 
$\theta_{\rm MC}$ is the approximation (used in {\tt CosmoMC}) to the angular size of the sound horizon at the time of last-scattering
$r_s(z_\ast)/D_A(z_\ast)$, $\tau$ is the Thomson scattering optical depth due to reionization, $n_s$ and $A_s$ are the spectral index and amplitude of the 
primordial curvature perturbations, respectively, and $r_{0.05}$ is the tensor-to-scalar ratio at $k_0=0.05$ Mpc$^{-1}$.
Other parameters, such as $\Omega_\Lambda$, $\Omega_m$, $\sigma_8$, $H_0$, $r_{0.002}$, and so on, are the derived parameters. 

In this base cosmology, there are three active neutrino species. Due to non-instantaneous decoupling corrections and other subtle corrections, the 
effective number of relativistic species in the standard cosmology is $N_{\rm eff}=3.046$. 
A minimal-mass normal hierarchy for the neutrino mass is assumed in the base cosmology, i.e., only one massive eigenstate with $m_\nu=0.06$ eV 
($\Omega_\nu h^2\approx \sum m_\nu/93.04~{\rm eV}\approx 0.0006$).

In this paper, we consider the extensions to this base cosmology. Neutrinos and extra relativistic components bring additional base parameters to the model.

\begin{itemize}
  \item Consider the total mass of active neutrinos. In this case, a degenerate model is assumed in which the three active neutrino species are 
  degenerate in mass and the total mass $\sum m_\nu$ is a free parameter. Thus, in this extension, one additional base parameter, $\sum m_\nu$, is introduced.
 
  \item Consider the extra neutrino-like radiation. 
  In this case, the extra relativistic degrees of freedom are effectively massless.
 The total mass of active neutrinos $\sum m_\nu$ is kept fixed at 0.06 eV, but the parameter $N_{\rm eff}$ is free.
 Thus, in this extension, one additional base parameter, $N_{\rm eff}$, is introduced.

 \item Simultaneously consider the active neutrino mass and extra radiation. In this case, the parameters $N_{\rm eff}$ and $\sum m_\nu$ are both free. 
 So, two additional parameters, $N_{\rm eff}$ and $\sum m_\nu$, are introduced.

 \item Consider the massive sterile neutrino. In this case, the total mass of active neutrinos $\sum m_\nu$ is kept fixed at 0.06 eV, but we add one massive 
 sterile neutrino in the model. Thus, two additional parameters, $N_{\rm eff}$ and $m_{\nu,{\rm sterile}}^{\rm eff}$, are introduced.

\end{itemize}

We use flat priors for the base parameters. When the base parameters are varied, the prior ranges are chosen to be much wider than the posterior so that 
the results of parameter estimation are not affected. The priors are set following the Planck Collaboration~\cite{planck}. In addition to these priors, a ``hard'' prior on the 
Hubble constant $H_0$ of [20, 100] km s$^{-1}$ Mpc$^{-1}$ is imposed.

\subsection{Observational data}

We consider the following data sets:

\begin{itemize}
  \item Planck+WP: the CMB TT angular power spectrum data from Planck~\cite{planck}, in combination with the large-scale 
  EE and TE polarization power spectrum data from 9-year WMAP~\cite{wmap9}.
  
  \item BAO: the latest measurement 
of the cosmic distance scale from the Data Release 11 (DR11) galaxy sample of
the Baryon Oscillation Spectroscopic Survey (BOSS) [that is part of the Sloan Digital Sky Survey III (SDSS-III)]: 
$D_V(0.32)(r_{d,{\rm fid}}/r_d)=(1264\pm 25)$~Mpc and $D_V(0.57)(r_{d,{\rm fid}}/r_d)=(2056\pm 20)$~Mpc, with 
$r_{d,{\rm fid}}=149.28$~Mpc~\cite{boss}.\footnote{There are also other BAO datasets, e.g., the 6dF with one point, 
$r_s/D_V(0.1)=0.336\pm 0.015$~\cite{6df}, and the WiggleZ with three points, 
$r_s/D_V(0.44)=0.0916\pm 0.0071$, $r_s/D_V(0.60)=0.0726\pm 0.0034$, and $r_s/D_V(0.73)=0.0592\pm 0.0032$~\cite{wigglez}. 
The inverse covariance matrix for the WiggleZ data is given in Ref.~\cite{wigglez} (see also Ref.~\cite{wmap9}). In this work, we choose to only use the 
latest two most accurate BAO measurements from BOSS DR11~\cite{boss}, which is sufficient for the purpose of this work in breaking the 
CMB parameter degeneracies.}

  \item $H_0$: the direct measurement of the Hubble constant using the cosmic distance ladder 
in the Hubble Space Telescope observations of Cepheid variables and type Ia supernovae, 
$H_0=(73.8\pm 2.4)~{\rm km}~{\rm s}^{-1}~{\rm Mpc}^{-1}$~\cite{h0}.\footnote{There is also another accurate $H_0$ measurement given by the {\it Carnegie Hubble Program}, 
$H_0=[74.3\pm 1.5~{\rm (statistical)}\pm 2.1~{\rm (systematic)}]$ km s$^{-1}$ Mpc $^{-1}$~\cite{Freedman2012}. 
This result agrees well with that of Ref.~\cite{h0}. 
In this work, we do not consider the measurement result of Ref.~\cite{Freedman2012}, but only use the one of Ref.~\cite{h0}.}

  \item SZ: the counts of rich clusters of galaxies from the sample of Planck thermal Sunyaev-Zeldovich (SZ) clusters constrain the combination of $\sigma_8$ and $\Omega_m$,  
$\sigma_8(\Omega_m/0.27)^{0.3}=0.78\pm 0.01$~\cite{tsz}.\footnote{The SZ cluster counts result quoted here is based on the use of the mass function from Ref.~\cite{Tinker2008}. 
A different mass function from Ref.~\cite{Watson2013} leads to a slightly different value of 
$\sigma_8(\Omega_m/0.27)^{0.3}=0.802\pm 0.014$. 
In addition, the result also depends, more or less, on the bias $(1-b)$ that is assumed to account for all possible 
observational biases including departure from hydrostatic equilibrium, absolute instrument calibration, temperature inhomogeneities, 
residual selection bias, etc. Numerical simulations based on the consideration of several ingredients of the gas physics of clusters 
give $(1-b)=0.8^{+0.2}_{-0.1}$. As pointed out by the Planck Collaboration~\cite{tsz}, adopting the central value, $(1-b)=0.8$, 
the constraints on $\Omega_m$ and $\sigma_8$ are in good agreement with previous measurements using clusters of galaxies. 
The result of $\sigma_8(\Omega_m/0.27)^{0.3}$ quoted here is derived by fixing $(1-b)=0.8$. If the bias $(1-b)$ is allowed to 
vary in the range [0.7,~1], the result is changed to $\sigma_8(\Omega_m/0.27)^{0.3}=0.764\pm 0.025$. 
Other values of $\sigma_8(\Omega_m/0.27)^{0.3}$ from various data combinations and analysis methods can be found in Table 2 of Ref.~\cite{tsz}.}

  \item Lensing: the CMB lensing power spectrum $C_\ell^{\phi\phi}$ from Planck~\cite{cmblensing},
  and also the combination of $\sigma_8$ and $\Omega_m$ given by the cosmic shear data of the weak lensing from the CFHTLenS survey, 
$\sigma_8(\Omega_m/0.27)^{0.46}=0.774\pm 0.040$~\cite{wl}.\footnote{We note that CMB lensing power spectrum and CFHTLens survey are two absolutely different, physically and observationally 
independent datasets.}
  
  \item BICEP2: the CMB angular power spectra (TT, TE, EE, and BB) data from BICEP2~\cite{bicep2}.

\end{itemize}

Actually, the Planck data are in tension with several astrophysical observations, as discussed by the Planck Collaboration~\cite{planck}, in the case of the 6-parameter base 
$\Lambda$CDM model. Planck data are in good agreement with the BAO data that are based on a simple geometrical measurement, so we can always combine Planck+WP 
with BAO without any question. But the Planck data are in tension with the $H_0$, SZ, and Lensing data. For the 6-parameter base $\Lambda$CDM model, 
the Planck+WP+highL data combination gives the fit results: $H_0=(67.3\pm 1.2)~{\rm km}~{\rm s}^{-1}~{\rm Mpc}^{-1}$, 
$\sigma_8(\Omega_m/0.27)^{0.3}=0.87\pm 0.02$, and $\sigma_8(\Omega_m/0.27)^{0.46}=0.89\pm 0.03$~\cite{planck}, which are in tension with the $H_0$ direct measurement~\cite{h0}, 
the cluster counts~\cite{tsz},\footnote{Actually, for the cluster counts, besides the Planck SZ-selected cluster sample~\cite{tsz}, there are also several other accurate datasets, 
including the SZ clusters from SPT~\cite{Reichardt:2012yj} and ACT~\cite{Hasselfield:2013wf}, and X-ray~\cite{Vikhlinin:2008ym} 
and optical richness~\cite{Rozo2010} selected cluster samples.
Constraints from these cluster samples on $\sigma_8(\Omega_m/0.27)^{0.3}$ can be found in Table 3 (and Fig.~10) of Ref.~\cite{tsz}. 
In this paper, we only focus on the Planck SZ cluster counts~\cite{tsz}.} 
and the cosmic shear measurement~\cite{wl} at the 2--3$\sigma$ level. In addition, Planck is also in mild tension with the 
SNLS type Ia supernova compilation (at about the 2$\sigma$ level).

Due to the complexity of these astrophysical data, these tensions can possibly be interpreted in terms of that some sources of systematic errors are not completely understood 
in these astrophysical measurement. An alternative explanation is that the base $\Lambda$CDM model is incorrect or should be extended.

The possibilities that the tensions between Planck and these astrophysical data might imply new physics have been explored. 
For example, the tension between Planck and the $H_0$ direct measurement might hint that dark energy is not the cosmological constant 
but is some dynamical field (or fluid). It is shown in Ref.~\cite{hde} that in a dynamical dark energy model, such as the constant $w$ model or the holographic dark energy model, the tension 
between Planck and $H_0$ is greatly reduced. But the mild tension between the Planck data and 
the SNLS type Ia supernova data may come from the systematic error, 
which could be greatly eliminated by considering the new effects of supernova,
such as the evolution of the 
color-luminosity parameter $\beta$, as analyzed in Refs.~\cite{beta1,beta2}. 

Sterile neutrino can also play a very significant role in relieving the tensions between Planck and the astrophysical observations. 
Involving sterile neutrino can increase the early-time Hubble expansion rate and the free-streaming damping, 
leading to the changes of the acoustic scale and the growth of cosmic structure, thus the 
tensions between Planck and $H_0$, cluster counts, and cosmic shear can simultaneously be greatly reduced when the massive sterile neutrino 
is considered~\cite{snu1,snu2,snu3}. 
Furthermore, very recently, it was shown that the tension between Planck and BICEP2 can also be significantly relieved when the sterile neutrino 
is involved in the model~\cite{zx14,WHu14}. 
Therefore, in the $\Lambda$CDM+$r$+$\nu_s$ model, almost all the tensions between Planck and other astrophysical observations can be simultaneously 
alleviated.

In this paper, we use the latest observational data to place constraints on the neutrino cosmological models. Since we use the uniform data sets, 
we actually can make a direct comparison for these models. We do not use the type Ia supernova data in this analysis because dynamical dark energy is 
not considered and also the systematic errors in the supernova data cannot be well quantified~\cite{beta1,beta2}. 
But we assume that other astrophysical data sets, such as $H_0$, SZ cluster counts, and cosmic shear, have accurately quantified estimates of systematic errors.
Since there is no tension between Planck and BAO, we can always safely use the Planck+WP+BAO data combination.
In order to measure the impacts from the other astrophysical observations on the neutrino physics, we can further combine the $H_0$+SZ+Lensing data in the analysis.
Furthermore, to see the role of the BICEP2 data play in constraining the neutrino cosmological models, we finally use an all data combination involving the BICEP2 data.
Thus, in our analysis, we use the data combinations: (i) Planck+WP+BAO, (ii) Planck+WP+BAO+$H_0$+SZ+Lensing, and (iii) Planck+WP+BAO+$H_0$+SZ+Lensing+BICEP2.
In the next section, we will report and discuss the fitting results of the neutrino cosmological models in the light of these data sets.



\section{Results and discussions}\label{sec:result}

For convenience, the four models considered in this paper are called: (i) $\Lambda$CDM+$r$+$\sum m_\nu$, 
(ii) $\Lambda$CDM+$r$+$N_{\rm eff}$, (iii) $\Lambda$CDM+$r$+$\sum m_\nu$+$N_{\rm eff}$, and 
(iv) $\Lambda$CDM+$r$+$N_{\rm eff}$+$m_{\nu,{\rm sterile}}^{\rm eff}$, respectively. 
The one- and two-dimensional joint, marginalized posterior 
distributions of the parameters for the four models are shown in Figs.~\ref{fig1}--\ref{fig4}. The grey, red, and blue contours (and curves) 
stand for the results of Planck+WP+BAO, Planck+WP+BAO+$H_0$+SZ+Lensing, and Planck+WP+BAO+$H_0$+SZ+Lensing+BICEP2 
data combinations, respectively. 
Detailed fit values for the cosmological parameters are given in Tables~\ref{tab1}--\ref{tab4}. 
In the tables, we quote the $\pm 1\sigma$ errors, but 
for the parameters that cannot be well constrained, we quote the 95\% CL upper limits.

\subsection{Constraints on the total mass of active neutrinos $\sum m_{\nu}$}

\begin{figure}[tbp]
\centering 
\includegraphics[scale=0.5]{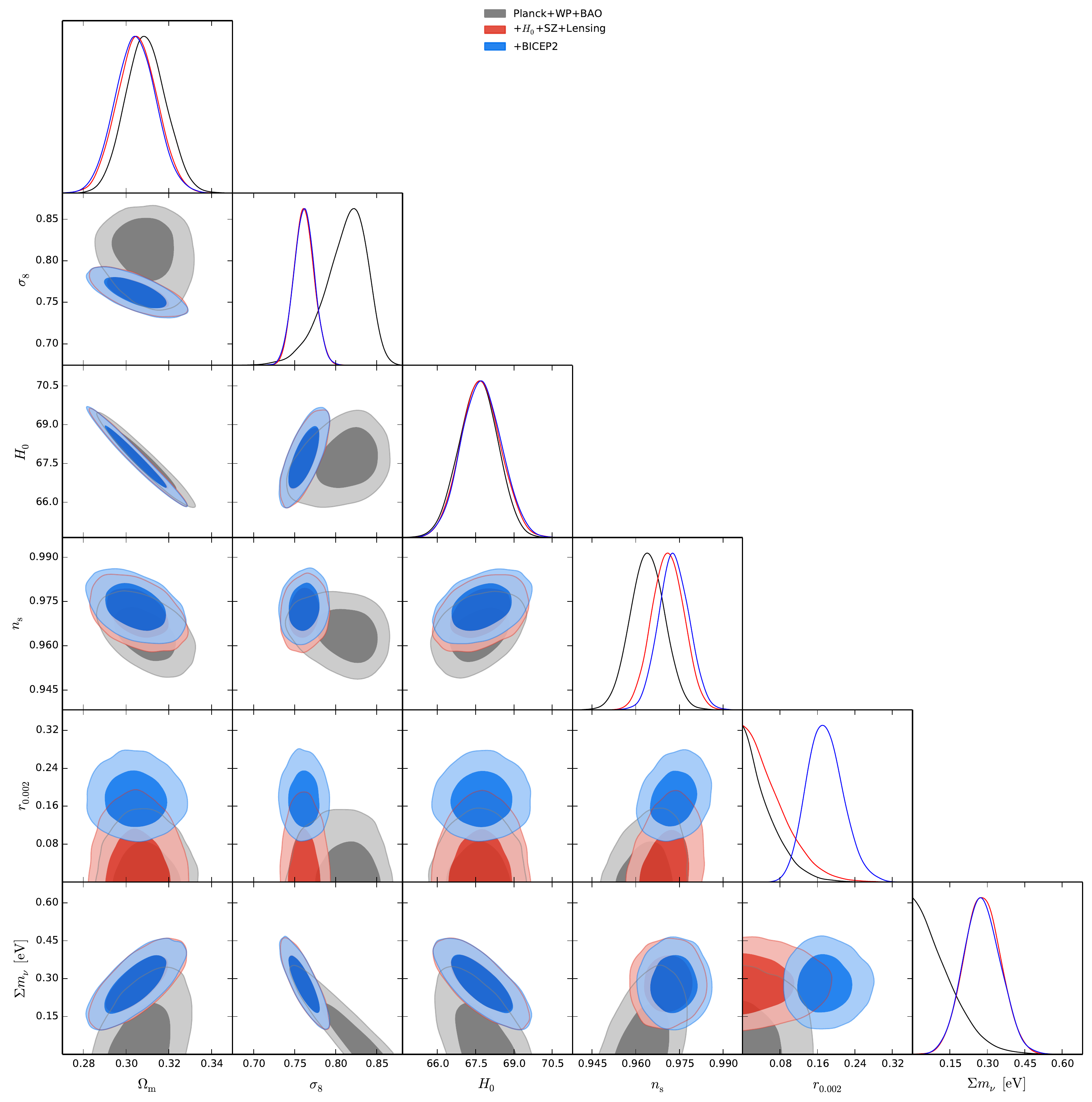}
\hfill
\caption{\label{fig1} Cosmological constraints on the $\Lambda$CDM+$r$+$\sum m_\nu$ model.}
\end{figure}

\begin{table}[tbp]
\centering
\begin{tabular}{|lccccccccc|}
\hline
 &\multicolumn{2}{c}{Planck+WP+BAO} & & \multicolumn{2}{c}{+$H_0$+SZ+Lensing} & & \multicolumn{2}{c}{+BICEP2} & \\
\cline{2-3}\cline{5-6}\cline{8-9}
Parameters & Best fit & 68\% limits && Best fit & 68\% limits && Best fit & 68\% limits & \\
\hline
$\Omega_bh^2$&$0.02223$&$0.02212\pm0.00024$&&$0.02249$&$0.0224\pm0.00024$&&$0.02237$&$0.02234^{+0.00023}_{-0.00025}$&\\
$\Omega_ch^2$&$0.1189$&$0.118^{+0.0019}_{-0.0017}$&&$0.1143$&$0.1145\pm0.0011$&&$0.1137$&$0.1143\pm0.0011$&\\
$100\theta_{\rm MC}$&$1.04185$&$1.04144\pm0.00055$&&$1.04152$&$1.04178\pm0.00053$&&$1.04191$&$1.04182^{+0.00054}_{-0.00060}$&\\
$\tau$&$0.095$&$0.091^{+0.012}_{-0.014}$&&$0.096$&$0.097^{+0.013}_{-0.014}$&&$0.107$&$0.096^{+0.013}_{-0.014}$&\\
$\sum m_\nu$&$0.00$&$<0.28$&&$0.249$&$0.280^{+0.072}_{-0.071}$&&$0.307$&$0.279^{+0.071}_{-0.078}$&\\
$n_s$&$0.9636$&$0.9641\pm0.0057$&&$0.9755$&$0.9711\pm0.0054$&&$0.9775$&$0.9732^{+0.0051}_{-0.0050}$&\\
${\rm{ln}}(10^{10} A_s)$&$3.098$&$3.087\pm0.025$&&$3.09$&$3.089^{+0.024}_{-0.027}$&&$3.111$&$3.088^{+0.024}_{-0.027}$&\\
$r_{0.05}$&$0.00$&$<0.13$&&$0.00$&$<0.16$&&$0.179$&$0.179^{+0.033}_{-0.038}$&\\
\hline
$\Omega_\Lambda$&$0.6966$&$0.6908^{+0.0099}_{-0.0091}$&&$0.6984$&$0.6947\pm0.0092$&&$0.6965$&$0.6955^{+0.0093}_{-0.0092}$&\\
$\Omega_m$&$0.3034$&$0.3092^{+0.0091}_{-0.0099}$&&$0.3016$&$0.3053\pm0.0092$&&$0.3035$&$0.3045^{+0.0092}_{-0.0093}$&\\
$\sigma_8$&$0.839$&$0.811^{+0.031}_{-0.018}$&&$0.768$&$0.762\pm0.012$&&$0.763$&$0.762\pm0.012$&\\
$H_0$&$68.26$&$67.61\pm0.74$&&$68.$&$67.71^{+0.75}_{-0.81}$&&$67.76$&$67.74^{+0.76}_{-0.82}$&\\
$r_{0.002}$&$0.00$&$<0.12$&&$0.00$&$<0.15$&&$0.178$&$0.177^{+0.034}_{-0.042}$&\\

\hline
$-\ln\mathcal{L}_{\rm{max}}$ &\multicolumn{2}{c}{4904.79} & & \multicolumn{2}{c}{4916.71} & & \multicolumn{2}{c}{4938.28} & \\
\hline
\end{tabular}
\caption{\label{tab1} Fitting results for the $\Lambda$CDM+$r$+$\sum m_\nu$ model. We quote $\pm 1\sigma$ errors, but 
for the parameters that cannot be well constrained, we quote the 95\% CL upper limits.}
\end{table}

Figure~\ref{fig1} and Table~\ref{tab1} summarize the fit results for the $\Lambda$CDM+$r$+$\sum m_\nu$ model.

From Fig.~\ref{fig1}, one can see that comparing to the Planck+WP+BAO data combination, the addition of the astrophysical data sets of 
$H_0$+SZ+Lensing impacts significantly on the constraint results of $\sigma_8$ and $\sum m_\nu$. 
But, $H_0$, $n_s$ and $r_{0.002}$ are not affected evidently.

The combination of Planck+WP+BAO gives $\sigma_8=0.811^{+0.031}_{-0.018}$, and when the data of $H_0$+SZ+Lensing are added, 
the fit result becomes  $\sigma_8=0.762\pm0.012$. 

Using the Planck+WP+BAO data cannot tightly constrain the neutrino mass, but can only obtain an upper limit 
$$\sum m_\nu<0.28~{\rm eV}\quad \mbox{(95\% CL; Planck+WP+BAO)}.$$
However, when the $H_0$+SZ+Lensing data are included, the neutrino mass can be tightly constrained,
$$\sum m_\nu=0.28\pm 0.07~{\rm eV}
\quad \mbox{(68\% CL;~Planck+WP+BAO+$H_0$+SZ+Lensing)}.$$
The posterior distribution is shown by the red curve in Fig.~\ref{fig1}.
Further including the BICEP2 data does not improve the constraint on the neutrino mass, 
$$\sum m_\nu=0.28^{+0.07}_{-0.08}~{\rm eV}
\quad \mbox{(68\% CL;~Planck+WP+BAO+$H_0$+SZ+Lensing+BICEP2)}.$$
The posterior distribution is shown in Fig.~\ref{fig1} by the blue curve which is nearly coincident with the red one. 
Thus, we find that in the $\Lambda$CDM+$r$+$\sum m_\nu$ model the combined cosmological data 
prefer a nonzero total mass of active neutrinos at about the 4$\sigma$ significance.

The BICEP2 does not affect other parameters, either, except for the tensor-to-scalar ratio $r$.
In the $\Lambda$CDM+$r$+$\sum m_\nu$ model, it is shown from Fig.~\ref{fig1} and Table~\ref{tab1} that 
the tension between Planck and BICEP2 cannot be effectively reduced. The Planck+WP+BAO data 
combination gives $r_{0.002}<0.12$ (95\% CL), and further adding $H_0$+SZ+Lensing data weakens the limit to 
$r_{0.002}<0.15$ (95\% CL). Including the BICEP2 data could improve the constraint on $r$ to 
$r_{0.002}=0.18^{+0.03}_{-0.04}$ (68\% CL).

\subsection{Constraints on the effective number of relativistic species $N_{\rm eff}$}

\begin{figure}[tbp]
\centering 
\includegraphics[scale=0.5]{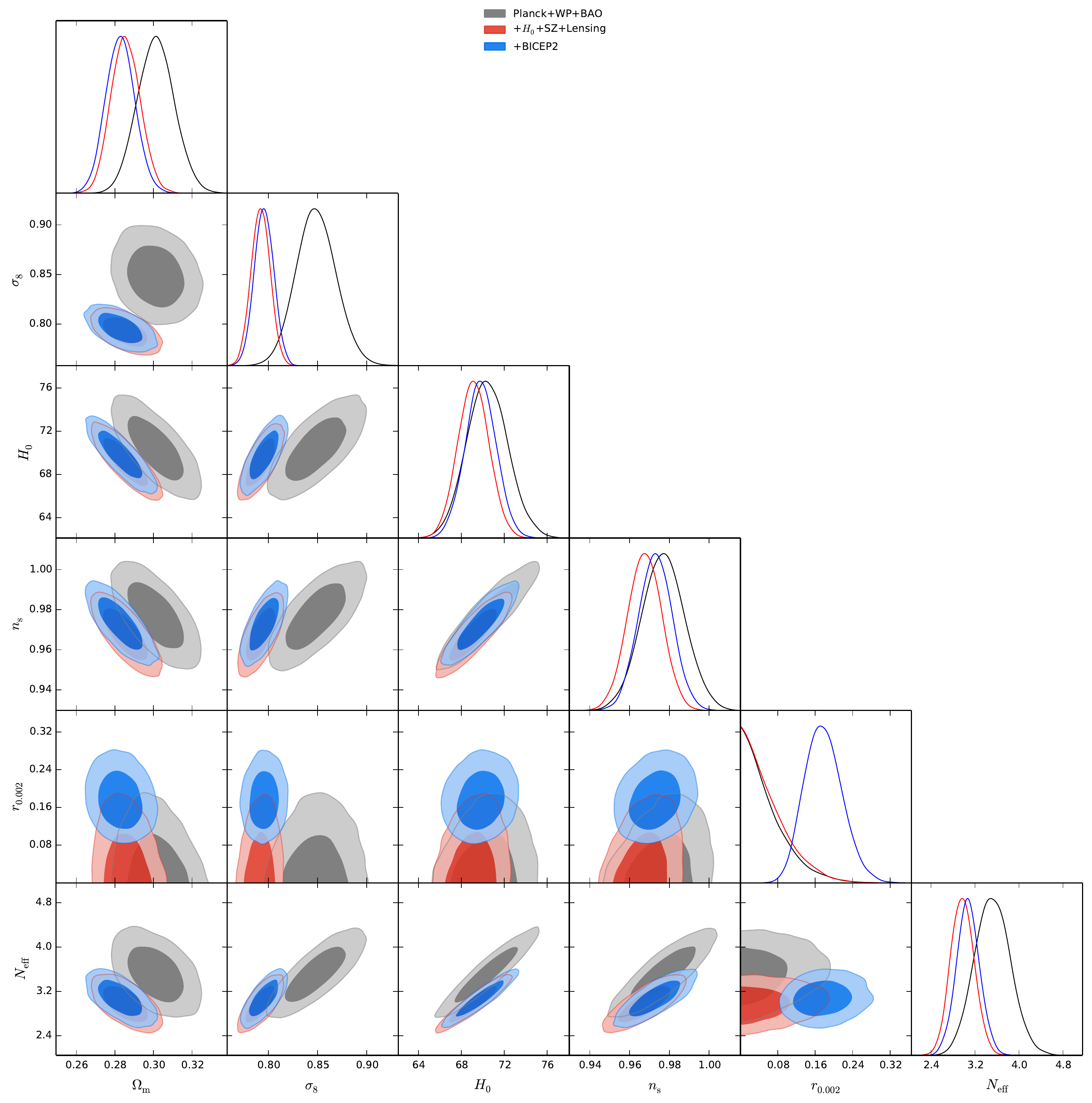}
\hfill
\caption{\label{fig2} Cosmological constraints on the $\Lambda$CDM+$r$+$N_{\rm eff}$ model.}
\end{figure}

\begin{table}[tbp]
\centering
\begin{tabular}{|lccccccccc|}
\hline
 &\multicolumn{2}{c}{Planck+WP+BAO} & & \multicolumn{2}{c}{+$H_0$+SZ+Lensing} & & \multicolumn{2}{c}{+BICEP2} & \\
\cline{2-3}\cline{5-6}\cline{8-9}
Parameters & Best fit & 68\% limits && Best fit & 68\% limits && Best fit & 68\% limits & \\
\hline
$\Omega_bh^2$&$0.02228$&$0.02234\pm0.00028$&&$0.02228$&$0.02238\pm0.00026$&&$0.02249$&$0.02238^{+0.00027}_{-0.00026}$&\\
$\Omega_ch^2$&$0.1274$&$0.1264\pm0.0055$&&$0.1135$&$0.1135^{+0.0031}_{-0.0035}$&&$0.1142$&$0.1148^{+0.0031}_{-0.0033}$&\\
$100\theta_{\rm MC}$&$1.04038$&$1.04061^{+0.00077}_{-0.00076}$&&$1.04209$&$1.04177^{+0.00068}_{-0.00067}$&&$1.04201$&$1.0416^{+0.00067}_{-0.00068}$&\\
$\tau$&$0.093$&$0.092^{+0.012}_{-0.014}$&&$0.073$&$0.076^{+0.010}_{-0.011}$&&$0.078$&$0.074^{+0.010}_{-0.011}$&\\
$N_{\rm eff}$&$3.55$&$3.52^{+0.31}_{-0.32}$&&$2.93$&$2.97^{+0.20}_{-0.22}$&&$3.07$&$3.07\pm0.20$&\\
$n_s$&$0.979$&$0.977\pm0.011$&&$0.9623$&$0.9676^{+0.0083}_{-0.0084}$&&$0.9785$&$0.973^{+0.0082}_{-0.0083}$&\\
${\rm{ln}}(10^{10} A_s)$&$3.113$&$3.109^{+0.028}_{-0.032}$&&$3.04$&$3.044\pm0.019$&&$3.054$&$3.046^{+0.019}_{-0.021}$&\\
$r_{0.05}$&$0.00$&$<0.15$&&$0.00$&$<0.16$&&$0.166$&$0.18^{+0.035}_{-0.039}$&\\
\hline
$\Omega_\Lambda$&$0.6959$&$0.6987^{+0.0093}_{-0.0092}$&&$0.711$&$0.7144^{+0.0074}_{-0.0073}$&&$0.7211$&$0.7171^{+0.0074}_{-0.0073}$&\\
$\Omega_m$&$0.3041$&$0.3013^{+0.0092}_{-0.0093}$&&$0.289$&$0.2856^{+0.0073}_{-0.0074}$&&$0.2789$&$0.2829^{+0.0073}_{-0.0074}$&\\
$\sigma_8$&$0.853$&$0.849\pm0.020$&&$0.7906$&$0.7922\pm0.0094$&&$0.7988$&$0.7956^{+0.0096}_{-0.0095}$&\\
$H_0$&$70.3$&$70.4^{+1.8}_{-1.9}$&&$68.7$&$69.1\pm1.4$&&$70.2$&$69.8\pm1.4$&\\
$r_{0.002}$&$0.00$&$<0.15$&&$0.00$&$<0.15$&&$0.166$&$0.178^{+0.036}_{-0.044}$&\\

\hline
$-\ln\mathcal{L}_{\rm{max}}$ &\multicolumn{2}{c}{4903.68} & & \multicolumn{2}{c}{4921.64} & & \multicolumn{2}{c}{4943.43} & \\
\hline
\end{tabular}
\caption{\label{tab2} Fitting results for the $\Lambda$CDM+$r$+$N_{\rm eff}$ model. We quote $\pm 1\sigma$ errors, but 
for the parameters that cannot be well constrained, we quote the 95\% CL upper limits.}
\end{table}

Figure~\ref{fig2} and Table~\ref{tab2} summarize the fit results for the 
$\Lambda$CDM+$r$+$N_{\rm eff}$ model.

The addition of the parameter $N_{\rm eff}$ can slightly relieve the tension between Planck and $H_0$.
The Planck+WP+BAO data combination gives $H_0=70.4^{+1.8}_{-1.9}$ km s$^{-1}$ Mpc$^{-1}$.
In the same case we also find a high amplitude for the present-day matter fluctuations, 
$\sigma_8=0.849\pm0.020$. When the $H_0$+SZ+Lensing data are added, the value of $H_0$ is not 
affected significantly, $H_0=69.1\pm 1.4$ km s$^{-1}$ Mpc$^{-1}$, but the value of $\sigma_8$ becomes 
much smaller, $\sigma_8=0.792\pm 0.009$ (with the error also shrinking significantly).

In the $\Lambda$CDM+$r$+$N_{\rm eff}$ model, the constraint results for the parameter $N_{\rm eff}$ 
are: $$N_{\rm eff}=3.52^{+0.31}_{-0.32}\quad \mbox{(68\% CL;~Planck+WP+BAO)},$$
$$N_{\rm eff}=2.97^{+0.20}_{-0.22}\quad \mbox{(68\% CL;~Planck+WP+BAO+$H_0$+SZ+Lensing)},$$
$$N_{\rm eff}=3.07\pm0.20\quad \mbox{(68\% CL;~Planck+WP+BAO+$H_0$+SZ+Lensing+BICEP2)},$$
which are all consistent with the standard value of 3.046.

We also find that in the $\Lambda$CDM+$r$+$N_{\rm eff}$ model the upper limit for the tensor-to-scalar ratio 
becomes a little bit higher, $r_{0.002}<0.15$, from the Planck+WP+BAO data, and this limit does not change 
when the $H_0$+SZ+Lensing data are added. So, this model cannot effectively alleviate the tension between 
Planck and BICEP2. When the BICEP2 data are included, the constraint on $r$ becomes 
$r_{0.002}=0.18\pm 0.04$.

The Planck+WP+BICEP2 constraints on the $\Lambda$CDM+$r$+$\sum m_\nu$ and $\Lambda$CDM+$r$+$N_{\rm eff}$ models 
were also discussed recently in Ref.~\cite{Giusarma:2014zza}.

\subsection{Simultaneous constraints on $N_{\rm eff}$ and $\sum m_{\nu}$}

\begin{figure}[tbp]
\centering 
\includegraphics[scale=0.45]{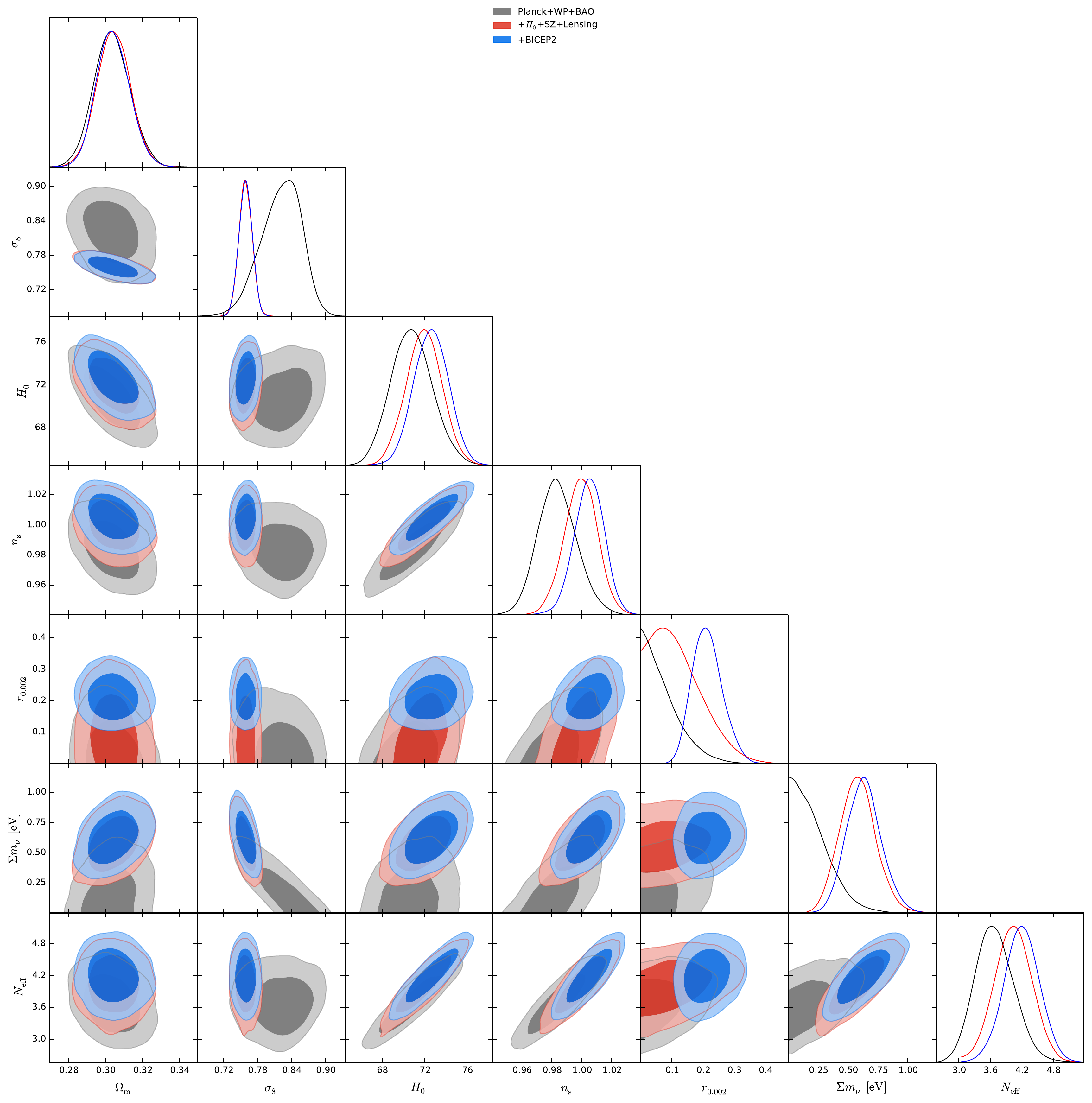}
\hfill
\caption{\label{fig3} Cosmological constraints on the $\Lambda$CDM+$r$+$\sum m_\nu$+$N_{\rm eff}$ model.}
\end{figure}

\begin{table}[tbp]
\centering
\begin{tabular}{|lccccccccc|}
\hline
 &\multicolumn{2}{c}{Planck+WP+BAO} & & \multicolumn{2}{c}{+$H_0$+SZ+Lensing} & & \multicolumn{2}{c}{+BICEP2} & \\
\cline{2-3}\cline{5-6}\cline{8-9}
Parameters & Best fit & 68\% limits && Best fit & 68\% limits && Best fit & 68\% limits & \\
\hline
$\Omega_bh^2$&$0.02215$&$0.02246^{+0.00029}_{-0.00034}$&&$0.02283$&$0.02290\pm0.00030$&&$0.02297$&$0.02294\pm0.00029$&\\
$\Omega_ch^2$&$0.1253$&$0.1275^{+0.0054}_{-0.0059}$&&$0.1232$&$0.1284\pm0.0050$&&$0.1307$&$0.1306\pm0.0047$&\\
$100\theta_{\rm MC}$&$1.041$&$1.04051\pm0.00072$&&$1.04115$&$1.04041^{+0.00068}_{-0.00069}$&&$1.04001$&$1.04022^{+0.00063}_{-0.00072}$&\\
$\tau$&$0.093$&$0.095^{+0.013}_{-0.015}$&&$0.109$&$0.108^{+0.014}_{-0.016}$&&$0.099$&$0.11\pm0.015$&\\
$\sum m_\nu$&$0.00$&$<0.50$&&$0.50$&$0.58^{+0.14}_{-0.15}$&&$0.62$&$0.63^{+0.13}_{-0.16}$&\\
$N_{\rm eff}$&$3.42$&$3.69^{+0.33}_{-0.40}$&&$3.66$&$4.04^{+0.35}_{-0.34}$&&$4.20$&$4.20\pm0.32$&\\
$n_s$&$0.972$&$0.984^{+0.012}_{-0.014}$&&$0.988$&$0.999\pm0.011$&&$1.0017$&$1.005^{+0.0104}_{-0.0093}$&\\
${\rm{ln}}(10^{10} A_s)$&$3.112$&$3.117^{+0.029}_{-0.034}$&&$3.134$&$3.144\pm0.031$&&$3.129$&$3.151\pm0.031$&\\
$r_{0.05}$&$0.00$&$<0.18$&&$0.022$&$<0.24$&&$0.21$&$0.195^{+0.036}_{-0.042}$&\\
\hline
$\Omega_\Lambda$&$0.6974$&$0.6965^{+0.0101}_{-0.0094}$&&$0.6927$&$0.6955^{+0.0088}_{-0.0086}$&&$0.6969$&$0.6958^{+0.0094}_{-0.0083}$&\\
$\Omega_m$&$0.3026$&$0.3035^{+0.0094}_{-0.0101}$&&$0.3073$&$0.3045^{+0.0086}_{-0.0088}$&&$0.3031$&$0.3042^{+0.0083}_{-0.0094}$&\\
$\sigma_8$&$0.856$&$0.821^{+0.041}_{-0.029}$&&$0.756$&$0.759\pm0.011$&&$0.751$&$0.759^{+0.012}_{-0.011}$&\\
$H_0$&$69.9$&$70.8^{+1.8}_{-2.1}$&&$70.2$&$71.9\pm1.6$&&$72.7$&$72.6\pm1.6$&\\
$r_{0.002}$&$0.00$&$<0.19$&&$0.021$&$<0.27$&&$0.23$&$0.215^{+0.041}_{-0.054}$&\\

\hline
$-\ln\mathcal{L}_{\rm{max}}$ &\multicolumn{2}{c}{4904.18} & & \multicolumn{2}{c}{4913.43} & & \multicolumn{2}{c}{4933.06} & \\
\hline
\end{tabular}
\caption{\label{tab3} Fitting results for the $\Lambda$CDM+$r$+$\sum m_\nu$+$N_{\rm eff}$ model. We quote $\pm 1\sigma$ errors, but 
for the parameters that cannot be well constrained, we quote the 95\% CL upper limits.}
\end{table}

Figure~\ref{fig3} and Table~\ref{tab3} summarize the fit results for the 
$\Lambda$CDM+$r$+$\sum m_\nu$+$N_{\rm eff}$ model.

In this model, the tension between Planck and $H_0$ direct measurement can be significantly reduced. 
The Planck+WP+BAO data combination gives $H_0=70.8^{+1.8}_{-2.1}$ km s$^{-1}$ Mpc$^{-1}$, which is 
improved to $H_0=71.9\pm 1.6$ km s$^{-1}$ Mpc$^{-1}$ when the $H_0$+SZ+Lensing data are included.
The Planck+WP+BAO data combination favors a high $\sigma_8$ value, $\sigma_8=0.821^{+0.041}_{-0.029}$, 
and the inclusion of the $H_0$+SZ+Lensing data improves the constraint to $\sigma_8=0.759\pm0.011$.
Further adding the BICEP2 data does not change these constraints evidently.

In the $\Lambda$CDM+$r$+$\sum m_\nu$+$N_{\rm eff}$ model, the constraint results for the parameters 
$N_{\rm eff}$ and $\sum m_{\nu}$ are:
$$
\left.
\begin{array}{c}
N_{\rm eff} = 3.69^{+0.33}_{-0.40}~~(68\%~{\rm CL}) \\
\sum m_\nu< 0.50~ {\rm eV}~~(95\%~{\rm CL})
\end{array}
\right\} \quad\mbox{(Planck+WP+BAO)},
$$
$$
\left.
\begin{array}{c}
N_{\rm eff} = 4.04^{+0.35}_{-0.34} \\
\sum m_\nu=0.58^{+0.14}_{-0.15}~{\rm eV}
\end{array}
\right\} \quad\mbox{(68\% CL;~Planck+WP+BAO+$H_0$+SZ+Lensing)},
$$
$$
\left.
\begin{array}{c}
N_{\rm eff} = 4.20\pm 0.32 \\
\sum m_\nu=0.63^{+0.13}_{-0.16}~{\rm eV}
\end{array}
\right\} \quad\mbox{(68\% CL;~Planck+WP+BAO+$H_0$+SZ+Lensing+BICEP2)}.
$$
We find that with the basic data combination Planck+WP+BAO, only an upper limit for the total mass of active 
neutrinos can be given, but the weak preference for $N_{\rm eff}>3.046$ at about the 1.6$\sigma$ level is shown.
Combining the $H_0$+SZ+Lensing data can tightly constrain both $\sum m_\nu$ and $N_{\rm eff}$, giving the 
evidence for nonzero mass of active neutrinos and $\Delta N_{\rm eff}\equiv N_{\rm eff}-3.046>0$ at the 3.9$\sigma$ 
and 2.9$\sigma$, respectively. Further adding the BICEP2 data can improve the results to some extent, favoring 
$\sum m_\nu>0$ and $\Delta N_{\rm eff}>0$ at the 4.0$\sigma$ and 3.6$\sigma$ levels, respectively.

It is interesting to compare the current results with those derived from data before Planck and BICEP2. 
For example, using the WMAP7+BAO+$H_0$+X-ray cluster data combination, Burenin obtained $\sum m_\nu=0.47\pm 0.16$ eV 
and $N_{\rm eff}=3.89\pm 0.39$~\cite{Burenin2013}, which indicates the 
detections of $\sum m_\nu>0$ and $\Delta N_{\rm eff}>0$ at the 2.9$\sigma$ and 2.2$\sigma$ levels, respectively.

It is also important to show that this model is very helpful in reconciling the tension between Planck and BICEP2.
With only the Planck+WP+BAO data, we find that the upper limit on the tensor-to-scalar ratio $r$ is weakened to 
$r_{0.002}<0.19$ (95\% CL). Once the $H_0$+SZ+Lensing data are included, the limit on $r$ is further weakened to 
$r_{0.002}<0.27$ (95\% CL), which is well compatible with the BICEP2 result, $r_{0.002}=0.20^{+0.07}_{-0.05}$~\cite{bicep2}.
Combining with the BICEP2 data, the $r$ constraint is tightened to $r_{0.002}=0.22^{+0.04}_{-0.05}$.
We also notice that due to the positive correlation between $n_s$ and $r$ (see the $n_s$--$r_{0.002}$ contours in grey and red in Fig.~\ref{fig3}), 
once the tensor-to-scalar ratio $r$ is increased, the scalar spectral index $n_s$ is also enlarged.
According to the fitting results, the exact scale-invariant perturbation spectrum cannot be excluded but actually is favored in this model.

\subsection{Constraints on massive sterile neutrino with $N_{\rm eff}$ and $m_{\nu,{\rm sterile}}^{\rm eff}$}

\begin{figure}[tbp]
\centering 
\includegraphics[scale=0.45]{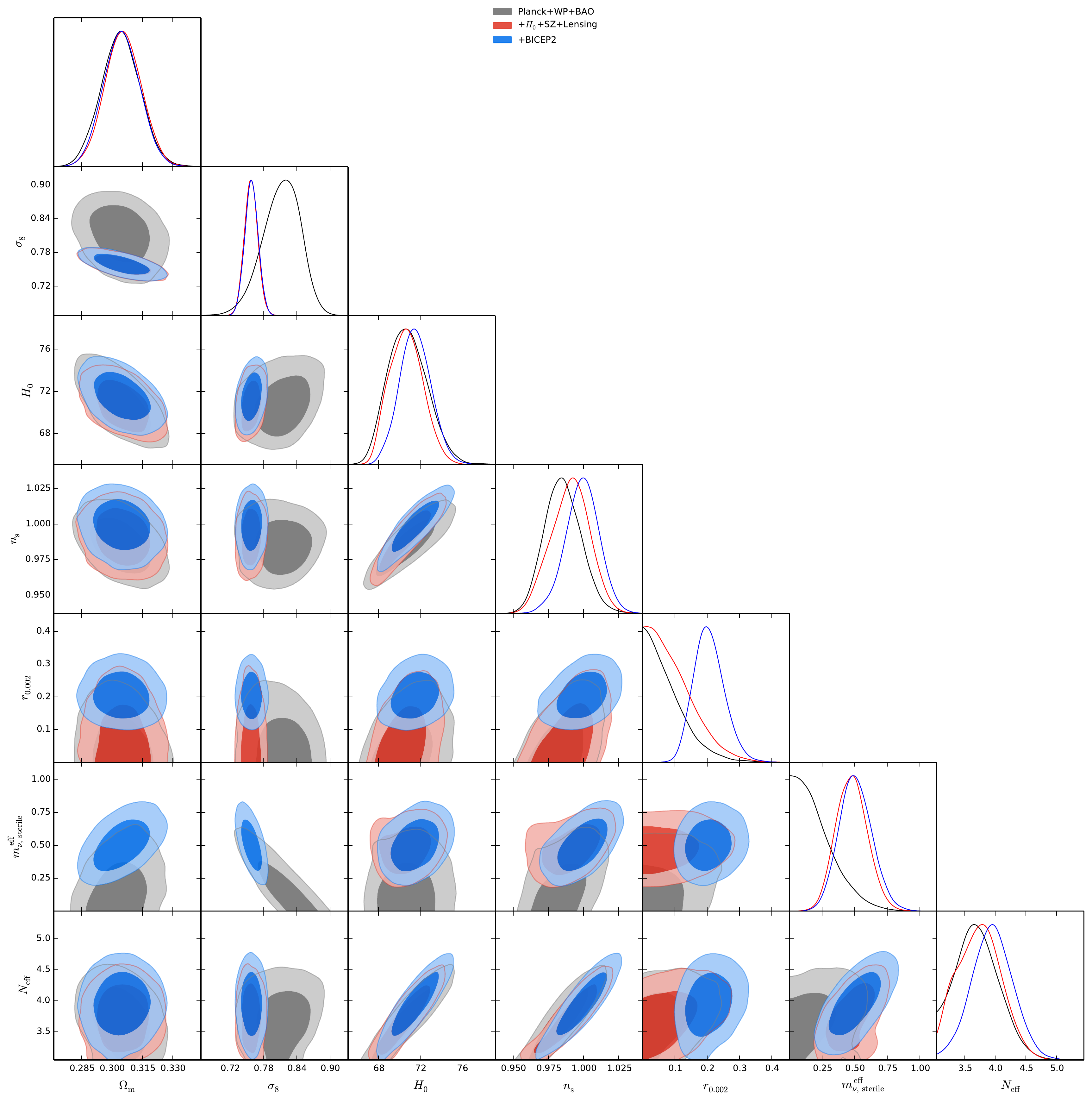}
\hfill
\caption{\label{fig4} Cosmological constraints on the $\Lambda$CDM+$r$+$N_{\rm eff}$+$m_{\nu,{\rm sterile}}^{\rm eff}$ model.}
\end{figure}

\begin{table}[tbp]
\centering
\begin{tabular}{|lccccccccc|}
\hline
 &\multicolumn{2}{c}{Planck+WP+BAO} & & \multicolumn{2}{c}{+$H_0$+SZ+Lensing} & & \multicolumn{2}{c}{+BICEP2} & \\
\cline{2-3}\cline{5-6}\cline{8-9}
Parameters & Best fit & 68\% limits && Best fit & 68\% limits && Best fit & 68\% limits & \\
\hline
$\Omega_bh^2$&$0.02229$&$0.0225\pm0.0003$&&$0.02261$&$0.02277^{+0.00027}_{-0.00028}$&&$0.02287$&$0.02282\pm0.00028$&\\
$\Omega_ch^2$&$0.1257$&$0.1273^{+0.0054}_{-0.0061}$&&$0.1168$&$0.1241^{+0.0052}_{-0.0056}$&&$0.1256$&$0.1271^{+0.0049}_{-0.0048}$&\\
$100\theta_{\rm MC}$&$1.04047$&$1.0405^{+0.00076}_{-0.00075}$&&$1.04162$&$1.04078\pm0.00074$&&$1.0411$&$1.0405\pm0.0007$&\\
$\tau$&$0.088$&$0.097^{+0.014}_{-0.015}$&&$0.101$&$0.106^{+0.014}_{-0.016}$&&$0.113$&$0.107^{+0.014}_{-0.016}$&\\
$m_{\nu,{\rm{sterile}}}^{\rm{eff}}$&$0.00$&$<0.51$&&$0.38$&$0.48^{+0.11}_{-0.13}$&&$0.51$&$0.51^{+0.12}_{-0.13}$&\\
$N_{\rm eff}$&$3.51$&$3.72^{+0.32}_{-0.40}$&&$3.28$&$3.75^{+0.34}_{-0.37}$&&$3.88$&$3.95\pm0.33$&\\
$n_s$&$0.974$&$0.985^{+0.012}_{-0.014}$&&$0.976$&$0.991^{+0.015}_{-0.013}$&&$0.998$&$0.999\pm0.011$&\\
${\rm{ln}}(10^{10} A_s)$&$3.1$&$3.12^{+0.030}_{-0.034}$&&$3.107$&$3.131^{+0.031}_{-0.035}$&&$3.147$&$3.14^{+0.031}_{-0.035}$&\\
$r_{0.05}$&$0.00$&$<0.19$&&$0.00$&$<0.21$&&$0.173$&$0.191^{+0.036}_{-0.041}$&\\
\hline
$\Omega_\Lambda$&$0.6998$&$0.6956\pm0.0093$&&$0.6984$&$0.6944^{+0.0087}_{-0.0088}$&&$0.6975$&$0.6952^{+0.0088}_{-0.0087}$&\\
$\Omega_m$&$0.3002$&$0.3044\pm0.0093$&&$0.3016$&$0.3056^{+0.0088}_{-0.0087}$&&$0.3025$&$0.3048^{+0.0087}_{-0.0088}$&\\
$\sigma_8$&$0.839$&$0.812^{+0.038}_{-0.029}$&&$0.758$&$0.758^{+0.011}_{-0.012}$&&$0.756$&$0.759\pm0.012$&\\
$H_0$&$70.4$&$70.8^{+1.7}_{-2.1}$&&$69.1$&$70.7^{+1.5}_{-1.8}$&&$71.5$&$71.5^{+1.4}_{-1.6}$&\\
$r_{0.002}$&$0.00$&$<0.20$&&$0.00$&$<0.23$&&$0.184$&$0.207^{+0.041}_{-0.052}$&\\

\hline
$-\ln\mathcal{L}_{\rm{max}}$ &\multicolumn{2}{c}{4904.07} & & \multicolumn{2}{c}{4913.24} & & \multicolumn{2}{c}{4933.82} & \\
\hline
\end{tabular}
\caption{\label{tab4} Fitting results for the $\Lambda$CDM+$r$+$N_{\rm eff}$+$m_{\nu,{\rm sterile}}^{\rm eff}$ model. We quote $\pm 1\sigma$ errors, but 
for the parameters that cannot be well constrained, we quote the 95\% CL upper limits.}
\end{table}

The $\Lambda$CDM+$r$+$N_{\rm eff}$+$m_{\nu,{\rm sterile}}^{\rm eff}$ model has been discussed in Refs.~\cite{zx14,WHu14}. 
In Ref.~\cite{zx14}, this model is also called $\Lambda$CDM+$r$+$\nu_s$ model, with $\nu_s$ denoting the sterile neutrino with two 
extra parameters $N_{\rm eff}$ and $m_{\nu,{\rm sterile}}^{\rm eff}$. 
In this paper, we duplicate the calculations in Ref.~\cite{zx14}, but we will provide more information about the fit results.
Figure~\ref{fig4} and Table~\ref{tab4} summarize the fit results for the 
$\Lambda$CDM+$r$+$N_{\rm eff}$+$m_{\nu,{\rm sterile}}^{\rm eff}$ model.

It has been discussed in Refs.~\cite{zx14,WHu14} (see also Refs.~\cite{snu1,snu2,snu3}) 
that the sterile neutrino can reconcile the tensions between Planck and 
other astrophysical observations such as the direct measurement of $H_0$~\cite{h0}, the Planck SZ cluster counts~\cite{tsz}, and the cosmic shear 
measurement~\cite{wl}. Here, we can see from Fig.~\ref{fig4} and Table~\ref{tab4} that the combination of Planck+WP+BAO gives 
$H_0=70.8^{+1.7}_{-2.1}$ km s$^{-1}$ Mpc$^{-1}$, and further combining with $H_0$, SZ, and Lensing data improves the result to 
$H_0= 70.7^{+1.5}_{-1.8}$ km s$^{-1}$ Mpc$^{-1}$. 
The Planck+WP+BAO data combination favors a high $\sigma_8$ value, $\sigma_8=0.812^{+0.038}_{-0.029}$, 
and the inclusion of the $H_0$, SZ, and Lensing data lowers the value to $\sigma_8=0.758^{+0.011}_{-0.012}$.
Further adding the BICEP2 data does not change these results evidently.

We now show the constraint results for the parameters $N_{\rm eff}$ and $m_{\nu,{\rm sterile}}^{\rm eff}$ in this model:
$$
\left.
\begin{array}{c}
N_{\rm eff} =3.72^{+0.32}_{-0.40}~~(68\%~{\rm CL}) \\
m_{\nu,{\rm sterile}}^{\rm eff}< 0.51~ {\rm eV}~~(95\%~{\rm CL})
\end{array}
\right\} \quad\mbox{(Planck+WP+BAO)},
$$
$$
\left.
\begin{array}{c}
N_{\rm eff} = 3.75^{+0.34}_{-0.37} \\
m_{\nu,{\rm sterile}}^{\rm eff}=0.48^{+0.11}_{-0.13}~{\rm eV}
\end{array}
\right\} \quad\mbox{(68\% CL;~Planck+WP+BAO+$H_0$+SZ+Lensing)},
$$
$$
\left.
\begin{array}{c}
N_{\rm eff} = 3.95\pm0.33 \\
m_{\nu,{\rm sterile}}^{\rm eff}=0.51^{+0.12}_{-0.13}~{\rm eV}
\end{array}
\right\} \quad\mbox{(68\% CL;~Planck+WP+BAO+$H_0$+SZ+Lensing+BICEP2)}.
$$
We find that the mass of sterile neutrino cannot be well constrained using only the basic data combination Planck+WP+BAO, but the 
addition of $H_0$, SZ, and Lending data significantly improves the constraint on the mass, strongly favoring a nonzero mass of sterile 
neutrino at the 3.6$\sigma$ statistical significance. The posterior distributions of $m_{\nu,{\rm sterile}}^{\rm eff}$ for the two cases are 
shown as grey and red curves, respectively, in Fig.~\ref{fig4}, and the direct comparison of the two curves is very impressive.
This shows that the SZ cluster data (as well as the $H_0$ and Lensing data) play an important role in constraining the mass of 
sterile neutrino, as discussed in Refs.~\cite{zx14,WHu14}. 
Further including the BICEP2 data improves the evidence for nonzero mass of sterile neutrino to be at the 3.9$\sigma$ significance.
For the $N_{\rm eff}$ constraints, the basic combination Planck+WP+BAO shows the preference for $\Delta N_{\rm eff}>0$ at the 1.7$\sigma$ 
level, and the inclusion of $H_0$+SZ+Lensing data improves slightly the preference for $\Delta N_{\rm eff}>0$ to be at the 1.9$\sigma$ level.
The BICEP2 data play a significant role in improving the constraint on $N_{\rm eff}$, which can be seen directly from the posterior distribution 
curves in Fig.~\ref{fig4}. Further adding the BICEP2 data favors the $\Delta N_{\rm eff}>0$ result at the 2.7$\sigma$ level.

The sterile neutrino can help reconcile the tension between Planck and BICEP2, as analyzed in Refs.~\cite{zx14,WHu14}. 
Using only Planck+WP+BAO can lead to $r_{0.002}<0.20$ (95\% CL), and including $H_0$+SZ+Lensing can give $r_{0.002}<0.23$ (95\% CL), 
consistent with the BICEP2 result. Further adding the BICEP2 data, we obtain the tightly constrained result, $r_{0.002}=0.21^{+0.04}_{-0.05}$.
As pointed out by Refs.~\cite{zx14,WHu14}, the increase of $r$ is at the expense of the increase of $n_s$, due to the positive correlation between 
$n_s$ and $r_{0.002}$ (as shown by the grey and red contours in the $n_s$--$r_{0.002}$ plane in Fig.~\ref{fig4}).
Hence, as the same as the $\Lambda$CDM+$r$+$\sum m_\nu$+$N_{\rm eff}$ model discussed in the last subsection, this model can resolve the 
tension between Planck and BICEP2, but at the same time cannot exclude the exact scale-invariant primordial perturbation spectrum.

The light massive sterile neutrino is motivated to explain the anomalies appearing in the 
short-baseline neutrino oscillation experiments~\cite{lsnd,miniboone,reactor,gallium,Giunti:2012tn,Giunti:2013aea}. 
It is of great interest to see that the evidence of the existence of the light sterile neutrino can be found in the existing cosmological data with high 
statistical significance (see also Refs.~\cite{zx14,WHu14,snu1,snu2,snu3}). Moreover, in this model almost all the tensions of Planck with 
other astrophysical observations can be simultaneously relieved.

The best-fit results, $\Delta N_{\rm eff}\approx 1$ and $m_{\rm sterile}^{\rm thermal}\approx m_{\nu,{\rm sterile}}^{\rm eff}\approx 0.5$ eV, 
derived in this paper and Refs.~\cite{zx14,WHu14},
indicate a fully thermalized sterile neutrino with sub-eV mass. However, the short baseline neutrino oscillation experiments 
prefer the mass of sterile neutrino at around 1 eV. So, there is still a tension on the sterile neutrino mass between the cosmological data and the 
short-baseline neutrino oscillation data. The implication of this tension for cosmology deserves further investigations. 
See Ref.~\cite{Archidiacono:2014apa} for a recent discussion.

\section{Conclusion}
\label{sec:concl}

After the detection of the PGWs by the BICEP2 experiment, the base standard cosmology should at least be 
extended to the 7-parameter $\Lambda$CDM+$r$ model. 
In this paper, we consider the extensions to this base $\Lambda$CDM+$r$ model by including additional 
base parameters relevant to neutrinos and/or other neutrino-like relativistic components. 
Four neutrino cosmological models are considered, i.e., the $\Lambda$CDM+$r$+$\sum m_\nu$, $\Lambda$CDM+$r$+$N_{\rm eff}$, $\Lambda$CDM+$r$+$\sum m_\nu$+$N_{\rm eff}$, and $\Lambda$CDM+$r$+$N_{\rm eff}$+$m_{\nu,{\rm sterile}}^{\rm eff}$ models. We use the current observational data to constrain these models.
The cosmological data used in this paper include: Planck+WP, BAO, $H_0$, Planck SZ 
cluster, Planck CMB lensing, cosmic shear, and  BICEP2 data. 
The main results of this paper are shown in Figs.~\ref{fig1}--\ref{fig4} and Tables~\ref{tab1}--\ref{tab4}.
Here, we summarize the findings from our analysis.

\begin{itemize}
\item  The $\Lambda$CDM+$r$+$\sum m_\nu$ model. With the Planck+WP+BAO data, we find a limit on 
the active neutrino mass, $\sum m_\nu<0.28~{\rm eV}$ (95\% CL). Including the $H_0$+SZ+Lensing data 
leads to a strikingly tight constraint: $\sum m_\nu=0.28\pm 0.07~{\rm eV}$, preferring a nonzero mass of 
active neutrinos at about the 4$\sigma$ level. Further adding the BICEP2 data does not improve the constraint 
on the mass. We also find that this model cannot alleviate the tension on $r$ between Planck and BICEP2.

\item The $\Lambda$CDM+$r$+$N_{\rm eff}$ model. 
Using only the Planck+WP+BAO data gives $N_{\rm eff}=3.52^{+0.31}_{-0.32}$, and further adding the 
$H_0$+SZ+Lensing data gives $N_{\rm eff}=2.97^{+0.20}_{-0.22}$, and 
combination of all data (including BICEP2) leads to $N_{\rm eff}=3.07\pm0.20$. 
These results are consistent with the standard value of 3.046.
We also find that this model cannot effectively alleviate the tension on $r$ between Planck and BICEP2.

\item The $\Lambda$CDM+$r$+$\sum m_\nu$+$N_{\rm eff}$ model. 
With the Planck+WP+BAO data, we obtain $\sum m_\nu< 0.50$ eV (95\% CL) and 
$N_{\rm eff} = 3.69^{+0.33}_{-0.40}$, so in this case only an upper limit on the total mass of active neutrinos 
can be given, but the weak preference for $N_{\rm eff}>3.046$ at about the 1.6$\sigma$ level is shown.
Combining with the $H_0$+SZ+Lensing data can lead to tight constraints, 
$\sum m_\nu=0.58^{+0.14}_{-0.15}$ eV and $N_{\rm eff} = 4.04^{+0.35}_{-0.34}$, 
giving the evidence for nonzero mass of active neutrinos and $\Delta N_{\rm eff}>0$ at the 3.9$\sigma$ 
and 2.9$\sigma$, respectively. Further adding the BICEP2 data can improve the results to 
$\sum m_\nu=0.63^{+0.13}_{-0.16}$ eV and $N_{\rm eff} = 4.20\pm 0.32$, 
favoring $\sum m_\nu>0$ and $\Delta N_{\rm eff}>0$ at the 4.0$\sigma$ and 3.6$\sigma$ levels, respectively.
We also show that this model is very helpful in relieving the tension between Planck and BICEP2. 
The increase of $r$ is at the cost of the increase of $n_s$, and consequently the exact scale-invariant spectrum cannot be excluded.

\item The $\Lambda$CDM+$r$+$N_{\rm eff}$+$m_{\nu,{\rm sterile}}^{\rm eff}$ model. 
With the Planck+WP+BAO data, we obtain $m_{\nu,{\rm sterile}}^{\rm eff}< 0.51$ eV (95\% CL) and 
$N_{\rm eff} =3.72^{+0.32}_{-0.40}$, thus in this case only an upper limit on the sterile neutrino mass can 
be derived and the preference for $\Delta N_{\rm eff}>0$ at the 1.7$\sigma$ level is shown.
Further including the $H_0$+SZ+Lensing data significantly improves the constraints, 
$m_{\nu,{\rm sterile}}^{\rm eff}=0.48^{+0.11}_{-0.13}$ eV and $N_{\rm eff} = 3.75^{+0.34}_{-0.37}$, 
favoring a nonzero mass of sterile neutrino and $\Delta N_{\rm eff}>0$ at the 3.6$\sigma$ and 1.9$\sigma$ 
levels, respectively. Finally, further adding the BICEP2 data improves the constraints to 
$m_{\nu,{\rm sterile}}^{\rm eff}=0.51^{+0.12}_{-0.13}$ eV and $N_{\rm eff} = 3.95\pm0.33$, 
showing the evidence of nonzero sterile neutrino mass and $\Delta N_{\rm eff}>0$ at the 
3.9$\sigma$ and 2.7$\sigma$ levels, respectively. 
It is shown that this model is very helpful in relieving the tension 
between Planck and BICEP2, and the expense of the increase of $r$ is the increase of $n_s$, thus 
the exact scale-invariant spectrum cannot be excluded in this case, either. 
The fitting results indicate a fully thermalized sterile neutrino with sub-eV mass, in tension with the 
short-baseline neutrino oscillation experiments that prefer the mass of sterile neutrino at around 1 eV. 
The implication of this tension for cosmology deserves further investigation.

\end{itemize}


\acknowledgments

We acknowledge the use of {\tt CosmoMC}.
This work was supported by the National Natural Science Foundation of
China (Grant No.~11175042) and the Fundamental Research Funds for the Central Universities (Grant No.~N120505003).



\end{document}